\documentclass[12pt]{article}%\documentstyle[12pt]{article}
\input epsf.sty
\usepackage{mathrsfs}
\usepackage{amsmath}
\usepackage{mathrsfs}
\usepackage{amssymb}
\usepackage{color}
\usepackage{psfrag}
\newcommand{\bea}{\begin{eqnarray}}
\newcommand{\eea}{\end{eqnarray}}
\newcommand{\be}{\begin{equation}}
\newcommand{\ee}{\end{equation}}
\newcommand{\vs}[1]{\vspace{#1 mm}}

\newcommand{\dsl}{\pa \kern-0.5em /}

\newcommand{\pa}{\partial}

\newcommand{\nn}{\nonumber\\}

\newcommand{\eqn}[1]{(\ref{#1})}

\begin{document}
\topmargin 0mm
\oddsidemargin 0mm

\begin{flushright}

USTC-ICTS-12-12\\

\end{flushright}

\vspace{2mm}

\begin{center}

{\Large \bf The phase structure of black D1/D5 (F/NS5) system
 \\in
canonical ensemble}

\vs{10}

{\large J. X. Lu\footnote{E-mail: jxlu@ustc.edu.cn}, Ran
Wei\footnote{E-mail: wei88ran@mail.ustc.edu.cn} and Jianfei
Xu\footnote{E-mail: jfxu06@mail.ustc.edu.cn}}

\vspace{4mm}

{\em
 Interdisciplinary Center for Theoretical Study\\
 University of Science and Technology of China, Hefei, Anhui
 230026, China\\}

\end{center}

\vs{10}

\begin{abstract}

In this paper, we explore means which can be used to change
qualitatively the phase structure of charged black systems. For
this, we consider a system of black D1/D5 (or its S-dual F/NS5). We
find that the  delocalized charged black D-strings (F-strings) alone
share the same phase structure as the charged black D5 branes
(NS5-branes), having no van der Waals-Maxwell liquid-gas type.
However, when the two are combined to form D1/D5 (F/NS5), the
resulting phase diagram has been changed dramatically to a richer
one, containing now the above liquid-gas type. The effect of adding
the charged D-strings (F-strings) on the phase structure can also be
effectively described as a slight increase of the transverse
dimensions to the original D5 (NS5). This may be viewed as a
connection between a brane charge and a fraction of spatial
dimension at least in a thermodynamical sense.
\end{abstract}

\newpage
\section{Introduction}
Recent studies \cite{Carlip:2003ne, Lundgren:2006kt} showed that a
large part of the phase structure of a black hole in asymptotically
anti-de Sitter (AdS) space \cite{Chamblin:1999tk, Chamblin:1999hg}
is not unique to the AdS black hole, but actually shared universally
by suitably stabilized black holes, say, in asymptotically flat
space, even in the presence of a charge $q$. For example, a
chargeless (suitably stabilized) black hole in asymptotically flat
space can also undergo a Hawking-Page transition at certain
temperature, now evaporating into a regular ``hot flat space"
instead of a regular ``hot empty AdS space" as for an AdS black
hole. Moreover, when $q \neq 0$, there exists also a critical charge
$q_c$ and for $q < q_c$, the phase diagram universally contains a
van der Waals-Maxwell liquid-gas type phase structure along with a
line of first-order phase transition terminating at a second-order
critical point with a universal critical exponent for specific heat
as\footnote{Using the standard definition for this exponent, we have
this value instead of $- 2/3$ as given in \cite{Carlip:2003ne,
Lu:2010xt}.} $2/3$.

Unlike an AdS black hole for which the asymptotically AdS space
itself acts as a reflecting box to stabilize the black hole, an
isolated asymptotically flat black hole is thermodynamically
unstable due to its Hawking radiation and needs to be stabilized
first before its phase structure can be analyzed properly. For this,
the standard practice is to place such a system inside a finite
spherical cavity \cite{York:1986it} with its surface temperature
fixed. In other words, a thermodynamical ensemble is considered
which can be either canonical or grand canonical, depending on
whether the charge inside the cavity or the potential at the surface
of the cavity is fixed \cite{Braden:1990hw}. Our focus in this paper
is the canonical ensemble, i.e., the charge inside the cavity is
fixed.

With the advent of AdS/CFT correspondence, the Hawking-Page
transition for AdS black hole `evaporating' into regular ``hot empty
AdS space" at certain temperature \cite{Hawking:1982dh} corresponds
to the confinement-deconfinement phase transition in large $N$ gauge
theory \cite{Witten:1998zw}. The very existence of the universal
phase structure mentioned above strongly suggests that this
universality is the result of the boundary condition rather than the
exact details of asymptotical metrics which can be either flat, AdS
or dS \cite{Carlip:2003ne, Lundgren:2006kt}. The boundary condition
realized in each case by the reflecting wall provides actually a
confinement to the underlying system. This may suggest that the AdS
holography is a result of such confinement rather than the detail
properties of AdS space and it naturally leads to the speculation
that a similar holography holds even in asymptotically flat space,
as hinted in \cite{Carlip:2003ne,Lundgren:2006kt}. If such a
holography exists indeed, the natural and interesting questions are:
what do the various thermal dynamical phase transitions correspond
to in the field theory defined on the underlying holographic screen?

The same phase structure in the chargeless case was also found
recently to be shared by the black p-branes in D-dimensional
spacetime with the brane worldvolume dimensions $d = 1 + p\,$ in
string/M theory and in the charged case the van der Waals-Maxwell
liquid-gas type phase structure holds also true when $\tilde d = D -
d - 2 > 2$ (note that $1 \le \tilde d \le 7$)\cite{Lu:2010xt,
Lu:2010au}. However, in the charged case we have a qualitative
different phase structure when $\tilde d < 2$ (there exists only the
$\tilde d = 1$ case) which actually resembles that of chargeless
case instead. The $\tilde d = 2$ case serves as a borderline in
phase structure which distinguishes the $\tilde d = 1$ case from the
$\tilde d > 2$ cases. For this case, there exists a `critical
charge' $q_c = 1/3$ and we have three subcases to consider,
depending on the charge $q > q_c, q = q_c$ and $q < q_c$ (like the
$\tilde d > 2$ cases), but we don't actually have a critical point
in the usual sense and for each subcase the phase structure looks
more like that of the $\tilde d = 1$ case (see \cite{Lu:2010xt} for
detail). So the qualitative phase structure in the charged case is
actually determined by $ \tilde d$, essentially the transverse
dimensions (actually $\tilde d + 2$) to the branes.

For $D = 10$, the $\tilde d = D - d - 2 = 2$ gives $d = 6$ and  the
corresponding black p-brane are either D5 or its S-dual NS5-branes.
The purpose of this paper is to explore means which can be used to
change qualitatively the phase structure of charged D5 or its S-dual
NS5-branes. We find that this can be achieved by adding delocalized
charged D-strings (or F-strings) to the black D5 (or NS5) branes.

Specifically, we consider to add the charged D1 (or F-strings) to
the D5 (or NS5) along one, say $x^1$, and delocalized along the
other four, say, $x^2, x^3, x^4, x^5$ of D5 (or NS5) worldvolume
spatial directions. In other words, we consider the system of D1/D5
(or its S-dual F/NS5). At first look, one may attribute the
qualitative change of phase structure of D5 (or NS5), which will be
described later in this paper, to the added D1 (or F-strings) since
one would naively think that for D-strings, the corresponding
$\tilde d = D - 2 -d = 10 - 2 -2 = 6 > 2$. This is actually not the
case. For the delocalized charged black D-strings (or F-strings),
their own phase structure (without the presence of D5 (or NS5)) is
actually the same as that of charged black D5 (or NS5) as we will
see in the later discussion. The surprise is that when both D5 (NS5)
and the delocalized D-strings (F-strings) are present, the resulting
phase structure is much richer and contains the van der
Waals-Maxwell liquid-gas type, dramatically different from the one
when only one type of branes are present. Concretely,  we have now
the reduced D5 (NS5) charge $q_5$ ($0 < q_5 < 1$) and the reduced D1
(F) charge $q_1$ ($0 < q_1 < 1$) and these two reduced charges form
a two-dimensional region enclosed by a square with unit side. As a
result, the present phase structure has a critical line instead of a
critical point as in the charged black p-brane case with $\tilde d >
2$. This critical line divides the two-dimensional $(q_1, q_5)$
charge region into two parts, with its one end approaching the point
$(0, 1/3)$ and the other end approaching the point $(1/3, 0)$. The
point $(0, 1/3)$ is on the $q_5$-axis (i.e., $q_1 = 0$) with $q_{5c}
= 1/3$, the black D5 brane `critical charge' mentioned earlier,
without the presence of delocalized D1. Similarly, the point $(1/3,
0)$ is on the $q_1$-axis with now $q_{1c} = 1/3$, the delocalized D1
`critical charge', without the presence of D5 branes. Any point
$(q_1, q_5)$ in the part, defined as the region enclosed by the line
interval $q_1 = 0, 0 \le q_5 \le 1/3$, the line interval $0 \le q_1
\le 1/3, q_5 =0$ and the critical line, gives a van der
Waals-Maxwell liquid-gas type phase structure, similar to the $q <
q_c$ case as in black p-branes with $\tilde d > 2$, for which the
equation of state can have three solutions if the cavity temperature
is chosen properly. Any point $(q_1, q_5)$ in the other part is
similar to the $q > q_c$ case as in black p-branes with $\tilde d >
2$, for which the equation of state has a unique solution. Any given
charge curve starting from one region, crossing the critical line
and entering the other region will give a phase structure similar to
the charged black p-branes with $\tilde d
> 2$ as discussed in \cite{Lu:2010xt}. So adding the delocalized
charged D-strings to D5 branes appears to increase effectively the
transverse dimensions to the D5 branes from $\tilde d = 2$ to
$\tilde d > 2$. Such an increase of $\tilde d$ is found to be
correlated, to certain extend, with that of the delocalized D-string
(F-string) charge\footnote{Actually it is related to the critical
charge of D-strings as we will discuss later.} and the effective
$\tilde d$ is actually slightly greater than $\tilde d = 2$  when
the delocalized D-string (F-string) charge is nonzero. So this
indicates a possible connection between the brane charge and the
effective spatial dimension, at least in the thermodynamical sense.

With one important exception, the other basic features of phase
structure found for D1/D5 are also common to the two remaining
systems D0/D4 and D2/D6 of the same type of D(p - 4)/Dp for $ p = 4,
6$, respectively. For either D0/D4 or D2/D6, this important
exception is that the underlying phase structure of D4 or D6 by
adding the delocalized D0 or D2 branes is, unlike the D1/D5 case,
not changed qualitatively\footnote{For D2/D6, with the addition of
 delocalized charged D2-branes, there is still some change of the
underlying phase structure and it is now similar to that of
D5-branes instead.}. For this reason, we will present the main
results for these two cases in the appendix.

    This paper is organized as follows. In section 2, we present the
 configuration of charged black D1/D5 in Euclidean signature and
derive the corresponding action from which we analyze the
thermodynamical stability and obtain the inverse of local
temperature for the purpose of understanding the underlying phase
structure. The details of the underlying phase structure analysis
are discussed in section 3. We explain what causes the dramatic
change of the phase structure of black D5 branes when delocalized
charged D-strings are added and discuss that the effect of adding
these D1 on phase structure can be effectively described by a system
of charged black p-branes with its $\tilde d$ slightly greater than
$2$. We conclude this paper in section 4. The remaining two cases
$D0/D4$ and $D2/D6$ of the same type of $D(p - 4)/Dp$ for $4 \le p
\le 6$ are discussed in the Appendix.

\section{The basic setup}
Our focus in this paper is to see how the added delocalized charged
D-strings to charged black D5-branes,  in the sense described in the
previous section, changes the phase structure of the
D5-branes\footnote{In the text, for simplicity, we  focus only on
the system of D1/D5 with the understanding that when the effective
string coupling becomes large, we should use its S-dual, i.e.,
F/NS5, instead, and the discussion remains exactly the same.}
qualitatively. For this purpose, we focus
 only on the usual related phases of the black D1/D5 system. For
simplicity, we will not consider here the corresponding bubble phase
as found relevant in \cite{Lu:2011da} in a similar fashion, which
can become globally stable when  certain conditions are met.

Let us consider the configuration of charged black D1/D5
\cite{Lu:2009tw}, now expressed in Euclidean signature as
 \bea \label{rv} ds^{2}
&=&{G_{5}}^{-\frac{1}{4}}{G_{1}}^{-\frac{3}{4}}\left(f dt^{2} + d
x_1^2\right) +
{G_{5}}^{-\frac{1}{4}}{G_{1}}^{\frac{1}{4}}\sum_{i=2}^{5} dx_i^{2} +
{G_{5}}^{\frac{3}{4}}{G_{1}}^{\frac{1}{4}} \left(\frac{d\rho^{2}}{f}
+ \rho^{2} d\Omega_{3}^{2} \right)\nn
A_{[2]}&=&-ie^{-\phi_{0}/2}\left[\tanh\theta_{1}-(1-
G_{1}^{-1})\coth\theta_{1}\right]dt\wedge dx^1\nn
A_{[6]}&=&-ie^{\phi_{0}/2}\left[\tanh\theta_{5}-(1-
{G_{5}}^{-1})\coth\theta_{5}\right]dt\wedge dx^{1}...\wedge
dx^{5}\nn e^{2(\phi-\phi_{0})}&=&G_{1}/ G_{5}, \eea where the metric
is in Einstein frame, each form field is obtained following
\cite{Braden:1990hw, Lu:2010xt} in such a way that the form field
vanishes at the horizon so that it is well defined in the local
inertial frame, $\phi_0$ is the asymptotic value of the dilaton,
${G}_{1,5}=1+\frac{\rho_{0}^{2}\sinh^{2}\theta_{1,5}}{\rho^{2}}$ and
$f=1-\frac{\rho_{0}^{2}}{\rho^{2}}$. The horizon is at $\rho =
\rho_0$ with a curvature singularity behind at $\rho = 0$. Note that
$\rho_0, \theta_1$ and $\theta_5$ are the parameters related to the
mass,  D1  charge and D5 charge of the system. For easily solving
the parameter constraints considered later, we make the following
radial coordinate transformation from $\rho$ to $r$, \be\label{vd}
\rho^2 = r^2 - r_-^2 , \,\,\, \rho_0^2 = r_+^2 - r_-^2,\ee with $r_+
\equiv \rho_0 \cosh\theta_5 \ge r_- \equiv \rho_0 \sinh\theta_5$.
Now the horizon is at $r = r_+$ with the curvature singularity at $r
= r_-$. We would like to stress that this particular choice of
radial coordinate is just for convenience and the final results are
independent of this. We can express the above configuration as \bea
\label{bd1d5} d s^2 &=& \triangle_-^{1/4} G^{- 3/4}_1 \left( \frac
{\triangle_+}{\triangle_-} d t^2 + d x_1^2\right) +
\triangle_-^{1/4} G_1^{1/4} \sum_{i = 2}^5 d x_i^2 +
\triangle_-^{1/4} G_1^{1/4} \left(\frac{d r^2}{\triangle_+
\triangle_-} + r^2 d \Omega_3^2\right),\nn
A_{[2]}&=&-ie^{-\phi_{0}/2}\left[\tanh\theta_{1}-(1-
G_{1}^{-1})\coth\theta_{1}\right]dt\wedge dx^1\nn
A_{[6]}&=&-ie^{\phi_{0}/2}\left[\tanh\theta_{5}-(1-
\triangle_-)\coth\theta_{5}\right]dt\wedge dx^{1}...\wedge dx^{5}\nn
e^{2(\phi-\phi_{0})}&=&G_{1} \triangle_-, \eea where we have now \be
\label{dgrelation} \triangle_\pm = 1 - \frac{r^2_\pm}{r^2},
\qquad\qquad G_1 = 1 + \left(1 -
\frac{\triangle_+}{\triangle_-}\right) \sinh^2 \theta_1.\ee Note
that to have a finite Euclidean action for the D1/D5 system, the
brane coordinates $x^i$ with $i = 1, \cdots 5$ should be compact.
For the above metric to be free from the conical singularity at the
horizon, the time coordinate `t' must be periodic with periodicity
\be \beta^* = 2 \pi r_+ \cosh\theta_1, \ee the inverse of
temperature of the black D1/D5 system at $r = \infty$. With this,
the inverse of local temperature at a given $r$ is \be
\label{localt} \beta = \triangle_+^{1/2} (\triangle_- G_1)^{- 3/8}
\beta^* = 2\pi \bar r_+ \triangle_+^{1/2} (\triangle_- G_1)^{- 1/2}
\cosh\theta_1,\ee where we have defined $\bar r_+ = r_+ (\triangle_-
G_1)^{1/8}$ for the reason given below.  Here $r$ is merely a
coordinate radius and the physical radius of the transverse 3-sphere
is from the above metric as $\bar r = r\, (\triangle_- G_1)^{1/8 }$
and the relevant local physical parameters $\bar r_\pm = r_\pm
(\triangle_- G_1)^{1/8}$. Note that the expression of
$\triangle_\pm$ remains the same even in terms of physical radial
coordinate $\bar r$ and parameters $\bar r_\pm$ as \be \triangle_\pm
= 1 - \frac{r^2_\pm}{r^2} = 1 - \frac{\bar r^2_\pm}{\bar r^2}.\ee

To study the equilibrium thermodynamics \cite{Gibbons:1976ue} in
canonical ensemble, the allowed configuration (the black D1/D5
system or its extremal one) should be placed in a cavity with a
fixed radius $\bar r_B$ ($> \bar r_+$) for the reason as explained
in the Introduction \cite{York:1986it}. The other fixed quantities
are the cavity temperature $1/\beta$, the physical periodicity of
each $x^i$ with $i = 1, \cdots 5$, the dilaton value $\bar \phi$ on
the surface of the cavity (at $\bar r = \bar r_B$) and the
charges/fluxes enclosed in the cavity $\bar Q_p$ ($p = 1, 5$),
respectively. In equilibrium, these fixed values are set equal to
the corresponding ones of the allowed configuration enclosed in the
cavity. For example, we set the charge \be\label{charge} \bar Q_p =
Q_p \equiv \frac{i}{\sqrt{2} \kappa}\int e^{-a(d)\phi} \ast F_{[d +
1]},\ee for $d = p + 1 = 2, 6$, respectively. In the above,  $\ast$
denotes the Hodge duality and the field strength $F_{[d + 1]}=  d
A_{[d]}$ with $A_{[d]}$ the corresponding form potential. With the
potentials $A_2, A_6$ given in \eqn{bd1d5}, we have \bea F_3 &=& - i
e^{- \phi_0/2} \frac{r (r_+^2 - r_-^2) \sinh 2\theta_1}{\left[r^2 -
r_-^2 + (r_+^2 - r_-^2)
\sinh^2 \theta_1\right]^2} d r\wedge dt \wedge d x^1, \\
 F_7 &=& - i e^{\phi_0/2} \frac{2 r_-^2}{r^3} \coth\theta_5 \, dr
 \wedge dt \wedge d x^1 \wedge \cdots \wedge d x^5.\eea
 We therefore have the D-string charge and D5-brane charge per unit five-brane volume, respectively, as
 \be\label{charge} Q_1 = \frac{\Omega_3 \bar V_4}{\sqrt{2} \kappa} e^{\bar\phi/2}
 \frac{(\bar r_+^2 - \bar r_-^2)\sinh 2\theta_1}{\triangle_-
 G_1},\qquad
Q_5 = \frac{\Omega_3}{\sqrt{2}\kappa} e^{- \bar\phi/2}\, 2\, \bar
r_+ \bar r_-,\ee where  $\Omega_n$ denotes the volume of a unit
$n$-sphere, $\kappa$ is a constant with $1/ (2 \kappa^2)$ appearing
in front of the Hilbert-Einstein action in canonical frame but
containing no asymptotic string coupling $g_s$. In the above, we
have expressed all the quantities in terms of their fixed (or
physical) correspondences, for example, we have replaced the
asymptotical string coupling $g_s$ in terms of the fixed effective
string coupling on the cavity via  $e^{\bar\phi} = e^{\phi(\bar
r_B)} \equiv g_s (\Delta_- G_1 )^{1/2}$.  The above physical volume
$\bar V_4 = (\triangle_- G_1)^{1/2} \, V_4^*$ with the coordinate
volume $V_4^* \equiv \int dx^2 dx^3 dx^4 d x^5$. In canonical
ensemble, it is the Helmholtz free energy which determines the
stability of equilibrium states and is related to the Euclidean
action by $F=I_E/\beta$ in the leading order approximation. So, in
order to understand the phase structure we will evaluate the action
for the black D1/D5.

The procedure for evaluating the Euclidean action of black
$p$-branes was given in detail in \cite{Lu:2010xt} following the
standard technique. The generalization to the present case is
straightforward, though the computation is a bit lengthy, by
considering one more piece contribution from the form field strength
$F_3$ and its potential $A_2$ (or $F_{p - 2}$ and its potential
$A_{p - 3}$ in general for $p \ge 4$) in addition to that from the
usual $F_7$ and its potential $A_6$ ($F_{p + 2}$ and its potential
$A_{p + 1}$) in the action. After a bit lengthy computations, we
have the Euclidean action for the black D1/D5 system as,
\bea\label{actiond1d5} I_E  &=& -\frac{\beta \bar V_5
\Omega_3}{\kappa^2}\bar r^2_B \left[2
\left(\frac{\Delta_+}{\Delta_-}\right)^{1/2} +
  \left(\Delta_+ \Delta_-\right)^{1/2} - 3 -
  \left(\frac{\Delta_+}{\Delta_-}\right)^{1/2} \left(1 - G_1^{-
  1}\right)\right]\nn
& & -\frac{2 \pi \bar r_+ \bar V_5 \Omega_3}{\kappa^2}\bar r_B^2
\left(1 - \frac{\triangle_+}{\triangle_-}\right)\left[1 + \frac{1 -
G_1^{- 1}}{\frac{\Delta_-}{\Delta_+} - 1}\right]^{1/2}, \eea with
$\Delta_\pm$, $G_1$ taking their respective value at $\bar r = \bar
r_B$. Since the Helmholtz free energy is given as $F = E - T S $, so
we have $I_E  = \beta E  - S$, where $E$ is the internal energy and
$S$ is the entropy of the black D1/D5. Thus we identify the internal
energy of the black D1/D5 on dividing the first term in
\eqn{actiond1d5} by $\beta$ and can be checked to match the ADM mass
per unit 5-brane volume of the system as $\bar r_B \to \infty$. The
second term in \eqn{actiond1d5} is the entropy of the system which
can be checked directly via $S = A/4 G$ by computing $A$, the
physical area transverse to the radial direction at $r = r_+$ from
the metric \eqn{bd1d5}, with $8 \pi G = \kappa^2$. In
\eqn{actiond1d5}, the 5-brane physical volume $\bar V_5 = G_1^{1/8}
\triangle_-^{5/8} V_5^*$ with the coordinate 5-brane volume $V_5^* =
\int d x^1 \cdots d x^5$. Note also that in obtaining action
\eqn{actiond1d5}, we have used the second expression in
\eqn{dgrelation} to replace the parameter $\theta_1$ in terms of
$\triangle_\pm$ and $G_1$ via (assuming $\theta_1 \ge 0$) \be
\label{thetaone}\sinh\theta_1 = \left(\frac{G_1 - 1}{1 -
\frac{\Delta_+}{\Delta_-}}\right)^{1/2}.\ee In action
\eqn{actiond1d5}, it appears that at this stage we have three
variables $\bar r_+$, $\bar r_-$ and $G_1$ since $\bar V_5, \beta$
and $\bar r_B$ are the fixed boundary data (Note that
$\triangle_\pm$ is a function of $\bar r_\pm$, respectively). This
is actually not the case since we have not used the two charge
expressions given in \eqn{charge} for which the two charges are set
to fixed in the canonical ensemble. So we are left with only one
variable and it is actually $\bar r_+$. Given the fixed boundary
data, we can define two new fixed charge quantities $\tilde Q_1$ and
$\tilde Q_5$, respectively, from \eqn{charge} as \bea
\label{chargec} \tilde Q_1 &=& \frac{\sqrt{2} \kappa \bar Q_1 e^{-
\bar \phi/2}}{2 \bar V_4 \bar r_B^2 \Omega_3} = \left(1 -
G_1^{-1}\right)^{\frac{1}{2}}\left[1 -
\frac{\triangle_+}{\triangle_-} +
\frac{\triangle_+}{\triangle_-}\left(1 -
G_1^{-1}\right)\right]^{\frac{1}{2}} < 1 \nn \tilde Q_5^2 &=&
\frac{\sqrt{2} \kappa \bar Q_5 e^{\bar\phi/2}}{2 \Omega} = \bar r_+
\bar r_-,\eea where we have set $Q_1 = \bar Q_1$ and $Q_5 = \bar
Q_5$ and in the second equality of the first equation we have used
\eqn{thetaone} to express $\theta_1$ in terms of $\triangle_\pm$ and
$G_1$. Note that the fixed $\tilde Q_1$ is dimensionless while the
$\tilde Q_5$ has a dimension of length square. From the second
equation, we have $\bar r_- = \tilde Q_5^2 /\bar r_+$ and so $\bar
r_-$ is expressed in terms of $\bar r_+$ (therefore $\triangle_-$ is
also expressed in $\bar r_+$). From the first equation, we can solve
a quadratic equation to find the relevant root \be \label{gone} 1 -
G_1^{-1} =
\frac{1}{2}\left[\sqrt{\left(\frac{\triangle_-}{\triangle_+} -
1\right)^2 + 4 \tilde Q_1^2 \frac{\triangle_-}{\triangle_+}} -
\left(\frac{\triangle_-}{\triangle_+} - 1\right)\right],\ee where we
have used the fact $1 - G_1^{ - 1} > 0$. With these, we can see that
the action \eqn{actiond1d5} is a function of variable $\bar r_+$
only.

For simplicity, we define the relevant reduced action \bea
\label{reducedaction}\bar I_E &\equiv& \frac{\kappa^2 I_E}{2 \pi\bar
V_5 \Omega_3 \bar r_B^3} \nn &=& - \frac{\beta}{2\pi \bar r_B}
\left[2 \left(\frac{\Delta_+}{\Delta_-}\right)^{1/2} +
  \left(\Delta_+ \Delta_-\right)^{1/2} - 3 -
  \left(\frac{\Delta_+}{\Delta_-}\right)^{1/2} \left(1 - G_1^{-
  1}\right)\right]\nn
& & - \frac{\bar r_+}{\bar r_B}  \left(1 -
\frac{\triangle_+}{\triangle_-}\right)\left[1 + \frac{1 - G_1^{-
1}}{\frac{\Delta_-}{\Delta_+} - 1}\right]^{1/2}. \eea We also define
the following reduced quantities \be \label{reducedv5} x \equiv
\left(\frac{\bar r_+}{\bar r_B}\right)^2 < 1,\qquad \bar b \equiv
\frac{\beta}{4\pi \bar r_B}, \qquad q_5 = \left(\frac{\tilde
Q_5}{\bar r_B}\right)^2 < x, \qquad q_1 = \tilde Q_1.\ee Note that
$0 < q_5 < 1$ (since $x < 1$) and $0 < q_1 < 1$ (since  $\tilde Q_1
< 1$ from the first equation in \eqn{chargec}). With these, we have
\be\label{delta} \triangle_+ = 1 - \frac{\bar r_+^2}{\bar r_B^2} = 1
- x, \qquad \triangle_- = 1 - \frac{\bar r_-^2}{\bar r_B^2} = 1 -
\frac{q_5^2}{x},\ee where for $\triangle_-$ we have used $\bar r_- =
\tilde Q_5^2/\bar r_+$. In terms of these reduced quantities, we
have the reduced action \eqn{reducedaction} as \bea
\label{reduceda}\bar I_E &=& - 2\,\bar b \left[2
\left(\frac{\Delta_+}{\Delta_-}\right)^{1/2} +
  \left(\Delta_+ \Delta_-\right)^{1/2} - 3 -
  \left(\frac{\Delta_+}{\Delta_-}\right)^{1/2} \left(1 - G_1^{-
  1}\right)\right]\nn
& & - x^{1/2}  \left(1 -
\frac{\triangle_+}{\triangle_-}\right)\left[1 + \frac{1 - G_1^{-
1}}{\frac{\Delta_-}{\Delta_+} - 1}\right]^{1/2}, \eea where \be
\frac{\triangle_+}{\triangle_-} =
\left(\frac{\triangle_-}{\triangle_+}\right)^{-1} = \frac{1 - x}{1 -
\frac{q_5^2}{x}},\ee and $1 - G_1^{-1}$ is given in terms of
$\triangle_-/\triangle_+$ via \eqn{gone} with now $\tilde Q_1 =
q_1$. So the above reduced action is only a function of variable $x$
with a few fixed parameters $\bar b, q_1, q_5$.

   The thermal equilibrium of black D1/D5 with the cavity can be
   determined, after a rather lengthy calculation,
    from \be \frac{d \bar I_E}{ d x} \sim (\bar b - b_{q_1,
   q_5} (x)) \Rightarrow 0,\ee
   to be
   \be\label{eos} \bar b = b_{q_1, q_5} (\bar x).\ee
   The local minima of $\bar
   I_E$, therefore the local stability,
   require
\be\label{feminima} \left.\frac{d^2 \bar I_E}{d x^2}\right|_{x =
\bar x} \sim -\frac{d b_{q_1, q_5} (\bar x)}{d \bar x} > 0,\ee
   which  in turn needs the negative slope of $b_{q_1, q_5} (\bar x)$ at the equilibrium.
   In the above, the inverse of the reduced
   temperature function \be \label{irt} b_{q_1, q_5} (x) =
   \frac{1}{2} x^{1/2}
   \left(\frac{\triangle_+}{\triangle_-}\right)^{1/2} \left[1 +
   \frac{1 - G_1^{-1}}{\frac{\triangle_-}{\triangle_+} -
   1}\right]^{1/2},\ee
   which agrees with the inverse of local temperature \eqn{localt} divided by $4\pi \bar r$ at
   $\bar r = \bar r_B$, therefore a consistent check.
This function is the basis for the analysis of the underlying phase
structure which we will perform in the following section.

\section{The analysis of phase structure}

  Before we discuss the surprise mentioned in the Introduction, we
  first examine a few things implied in the function $b_{q_1, q_5}
  (x)$ derived in the previous section. First, when we set the
  delocalized D-string charge $q_1 = 0$, we end up with the inverse
  of the reduced temperature function $b_{q_5} (x)$, as expected, for the black 5-branes  given in
\cite{Lu:2010xt}
\be \label{5-brane} b_{0, q_5} (x) = b_{q_5} (x) =
\frac{1}{2} x^{1/2} \left(\frac{1 - x}{1 -
\frac{q_5^2}{x}}\right)^{1/2}.\ee
  Note here $q_5 < x < 1$. Let us examine what happens if we set the 5-brane
  charge $q_5 = 0$. We now have from \eqn{irt}
  \be\label{lstring} b_{q_1, 0} (x) = \frac{1}{2} (1 - x)^{1/2} \left(\frac{x + \sqrt{x^2 + 4
  q _1^2 (1 - x)}}{2}\right)^{1/2},\ee where we have used
  $\triangle_- = 1$ and the explicit expressions for $\triangle_+$ in \eqn{delta} and
  $1 - G_1^{-1}$ as given in \eqn{gone}, respectively. Note now $0 <
  x < 1$ and it appears that the above $b_{q_1, 0} (x)$ looks quite
  different from $b_{0, q_5} (x)$ given in \eqn{5-brane}. This
  apparent difference is actually due to the improper use of the
  variable $x$ for the present case. As we stress in the
  previous section, we make a special choice of the radial coordinate as given in \eqn{vd} as well as $r_\pm$ so that
  the charge constraints given in \eqn{chargec} are simplified greatly and can
  be solved explicitly. While this is good for convenience, it is
  not essential and our choice there prefers actually to the 5-branes.
  When $q_5 = 0$, we have the range of variable $x$ in $0 < x < 1$ which merely reflects the
  5-branes being chargeless, certainly not a good one in describing
  the delocalized strings. We expect a relevant variable $y$ in the
  range $q_1 < y < 1$ and this can be achieved via\footnote{One can
  show directly that in analogous to the variable $x$ favoring
  D5-branes, the variable $y$ is indeed the one favoring the
  delocalized D-strings in a similar fashion. For this, in analogous to $r$ and $r_\pm$ in \eqn{vd},
  we define here $\rho^2 = \tilde r^2 - \tilde r_-^2, \rho_0^2 =
  \tilde r_+^2 - \tilde r_-^2$ with $\tilde r_+ = \rho_0
  \cosh\theta_1$ and $\tilde r_- = \rho_0 \sinh\theta_1$. So we expect $y = (\tilde r_+/\tilde r_B)^2 = \rho^2_0 \cosh^2\theta_1 /(\rho^2_B + \rho^2_0 \sinh^2\theta_1)
  = x \cosh^2 \theta_1/(1 + x \sinh^2\theta_1)$ where $x = (\rho_0/\rho_B)^2$ when $q_5 = 0$. When $q_5 = 0$, $r_- = 0$
  and from the first equation in \eqn{charge} and the definition for $q_1$ given in \eqn{chargec} and \eqn{reducedv5}, we have
  $q_1  = x \cosh\theta_1 \sinh\theta_1/(1 + x
  \sinh^2\theta_1)$. Combining these two equations, we have $y^2 - x y - q_1^2 (1 - x) = 0$ whose proper solution is nothing but \eqn{yv} with $q_1 < y < 1$.}
  \be \label{yv} y = \frac{x + \sqrt{x^2 + 4 q_1^2 (1 - x)}}{2},
  \ee from which we solve for $1 - x$ as \be 1 - x = \frac{1 -
  y} {1 - \frac{q_1^2}{y}}.\ee Using the $y$ variable in
  \eqn{lstring}, we have \be b_{q_1, 0} (y) = \frac{1}{2} y^{1/2}\left(\frac{1 - y}{1 -
  \frac{q_1^2}{y}}\right)^{1/2},\ee which is identical in form to
  $b_{0, q_5} (x)$ given in \eqn{5-brane} but now with variable $y$ and its range $q_1 < y < 1$. So the delocalized charged black D-strings, unlike the localized charge black D-strings, have
  the same phase structure as the black D5-branes, described in detail in
  \cite{Lu:2010xt}. In other words, so long the phase structure is
  concerned, the delocalized charged black D-strings look no different from the
  black D5-branes, which is completely determined by the function $b_{q_1, 0} (y)$ or $b_{0,
  q_5} (x)$.

  Given what has been said above, one might naively conclude that when both the 5-branes and the
  delocalized strings are present, the underlying phase structure
  would remain the same as that when either are present. This is
  actually not the case, a surprise, as we will see in what follows.
  For this, let us examine the full expression $b_{q_1, q_5}
  (x)$ \eqn{irt} when both $q_1$ and $q_5$ are non-zero. The above
  discussion indicates already that the variable $x$ is not a good one
  since it favors 5-branes over the delocalized strings and we
  would like to have one which can give a symmetric representation
  between them. This can be achieved via the following\footnote{This
  variable $f$, in terms of the original variable $\rho$ (and $\rho_0$) given in \eqn{rv}, is $f = 1 -
  \rho_0^2/\rho_B^2$.}
  \be \label{fvariable} f = \frac{\triangle_+}{\triangle_-} = \frac{1 - x}{1 -
  \frac{q_5^2}{x}} < 1,\ee
  from which we can solve to give (noting $q_5 < x < 1$)
  \be \label{variablet} x = \frac{1 - f + \sqrt{(1 - f)^2 + 4 q_5^2 f}}{2}.\ee In the
  above,  note that $x = q_5$ corresponds to $f = 1$ while $x = 1$ to $f = 0$
  and the range for $f$ is now $0 < f < 1$. With the new variable $f$, we
  have the reduced action \eqn{reduceda}
  \bea \label{ractionf}\bar I_E &=& - 2 \bar b \left[2 f^{\frac{1}{2}} + f^{-\frac{1}{2}} - 3 - \frac{\sqrt{(1 - f)^2 + 4 q_1^2 f} + \sqrt{(1 - f)^2 + 4 q_5^2 f}}{2 f^{1/2}}\right]\nn
  &\,& - (1 - f)^{\frac{1}{2}} \left[\frac{1 - f + \sqrt{(1 - f)^2 + 4 q_1^2
  f}}{2}\right]^{\frac{1}{2}} \left[\frac{1 - f + \sqrt{(1 - f)^2 + 4 q_5^2
  f}}{2}\right]^{\frac{1}{2}},\eea and the function \eqn{irt}
  \be \label{bf} b_{q_1, q_5} (f) = \frac{1}{4} \left[\frac{f \left(1 - f + \sqrt{(1
  - f)^2 + 4 q_1^2 f}\right)\left(1 - f + \sqrt{(1 - f)^2 + 4 q_5^2 f}\right)}{1 -
  f}\right]^{1/2},\ee both of which do reflect the symmetry between $q_1$
  and $q_5$. The equation of state \eqn{eos} becomes now
  \be \label{eosf} \bar b = b_{q_1, q_5} (\bar f).\ee  Let us examine the behavior of $b_{q_1, q_5} (f)$ which
  will determine the phase structure of D1/D5 system. Note that
  $b_{q _1, q_5} (f) > 0$ for $0 < f < 1$ and when neither charge is
  zero,  we have\footnote{This same behavior can be seen from the
  original $b_{q_1, q_5} (x)$ \eqn{irt} using variable $x$: $b_{q_1, q_5} (x
  \rightarrow q_5) \rightarrow \infty$,  noting $\triangle_+/\triangle_- \rightarrow 1$ when $x \rightarrow q_5$, and $b_{q_1, q_5} (x
  \rightarrow 1) \rightarrow 0$.}
  \be \label{bbvsf} b_{q_1, q_2} (f \rightarrow 1) \rightarrow \infty, \qquad
  b_{q_1, q_5} (f \rightarrow 0) \rightarrow 0.\ee The corresponding function
  for black p-branes  with $\tilde d > 2$ has also the same behavior\cite{Lu:2010xt}, critical to the underlying phase structure. So we expect
  here the similar phase structure, namely, the existence of van der
  Waals-Maxwell liquid-gas type phase structure\footnote{Note that from \eqn{variablet}, we have
   \be 2\, d x = - \left[1 + \frac{1 - f - 2 q_5^2}{\sqrt{(1 - f)^2 + 4 q_5^2 f}}\right]
   \, d f, \ee and since $\sqrt{(1 - f)^2 + 4 q_5^2 f} > |1 - f - 2
   q_5^2|$, we have $d x \sim - d f$ and hence the negative slope of $b_{q_1, q_5} (x)$, determining the local minima of the free energy
   \eqn{feminima} at an equilibrium point,
    corresponds to the positive slope of $b_{q_1, q_5} (f)$. So in
    what follows, we seek the positive slope of $b_{q_1, q_5} (f)$
    for the local minima of free energy at an equilibrium.}.
   Unlike the black p-brane case, we have
  here a critical line instead, which can be determined from the following two conditions:
  \be \label{criticalcd} \frac{d\, b_{q_1, q_5} (f)}{d \,f\,\,\,\,} = 0, \qquad \frac{ d^2\, b_{q_1,
  q_5} (f)}{d \,f^2\,\,\,\,} = 0.\ee From the above, we have
  \be \label{cc} \frac{1}{A} + \frac{1}{B} = \frac{2 (2 - f)}{1 + f}, \qquad
  \frac{1}{A^3} + \frac{1}{B^3} = \frac{2 (2 - 3 f + 6 f^2 -
  f^3)}{(1  + f)^3},\ee
  where \be\label{ab} A = \sqrt{1 + 4 q_1^2 \frac{f}{(1 - f)^2}},\qquad B =
  \sqrt{1 + 4 q_5^2 \frac{f}{(1 - f)^2}}.\ee
 Unlike the black p-brane case\cite{Lu:2010xt}, we have here three critical quantities $q_{1c}, q_{5c}, f_c$ to be determined but with only two equations in
 \eqn{cc}. This must imply that we will have a critical line rather than a critical point as for the black p-brane case. The critical charges $q_{5c}$ and
  $q_{1c}$, representing a line, can be solved in terms of the critical
parameter $f_c$ as
 \bea \label{criticalc} q_{5c} &=& \sqrt{\frac{(1 + f_c)^2
\left[\sqrt{(2 - f_c)(3 f_c -2)} \pm (2 - f_c)^2\right]^2}{4 f_c (10
- 5 f_c + f_c^2)^2} - \frac{(1 - f_c)^2}{4 f_c}},\nn q_{1c} &=&
\sqrt{\frac{(1 + f_c)^2 \left[\sqrt{(2 - f_c)(3 f_c -2)} \mp (2 -
f_c)^2\right]^2}{4 f_c (10 - 5 f_c + f_c^2)^2} - \frac{(1 -
f_c)^2}{4 f_c}}.\eea In the above, the existence of the critical
line actually requires $f_c \ge 2/3$. So the range for $f_c$ is $2/3
\le f_c < 1$.
\begin{figure}
\psfrag{A}{$q_1$} \psfrag{B}{$q_5$} \psfrag{C}{$(1/3, 0)$}
\psfrag{D}{$(0, 1/3)$} \psfrag{E}{Symmetric point
($\frac{\sqrt{6}}{16}, \frac{\sqrt{6}}{16}$)} \psfrag{F}{Critical
line}
\begin{center}
\includegraphics{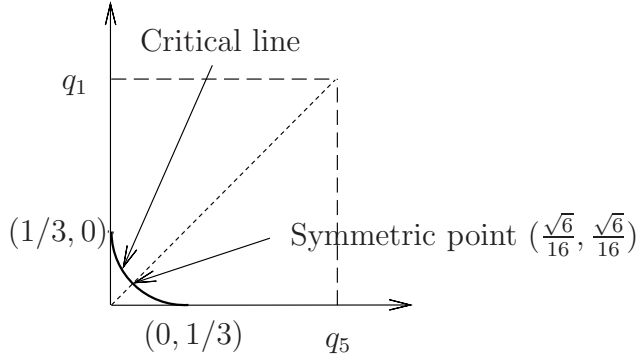}
\end{center}
  \caption{The reduced charges $q_1$ and $q_5$ form a two-dimensional
  region bounded by a square with a unit side and the critical line divides
the region with the upper part giving rise to the phase similar to
the $q > q_c$ case for black p-branes when $\tilde d > 2$ while the
lower part to the $q < q_c$ case containing a van der Waals-Maxwell
liquid-gas type structure.}
\end{figure}

Given $0 < q_1 < 1$ and $0 < q_5 < 1$, the two reduced charges $q_1$
and $q_5$ form a two dimensional region enclosed by a square with a
unit side as shown in Fig. 1. The critical line divides the region
into two parts, the upper part and the lower one. Given what we know
about the phase structure of black 5-branes (corresponding to the
$q_5$-axis with $q_1 = 0, 0 < q_5 < 1$ and $q_{5c} = 1/3$) or the
delocalized D-strings (corresponding to the $q_1$-axis with $0 < q_1
< 1, q_5 = 0$ and $q_{1c} = 1/3$), we should have the upper part of
the region to give rise to the phase structure similar to the $q >
q_c$ case for black p-branes when $\tilde d > 2$ and the lower part
to the $q < q_c$ case containing the van der Waals-Maxwell
liquid-gas type structure. Before we examine this further, we would
like to discuss a bit more about the charge parameter-space
structure.

Given that the delocalized charged black strings and the charged
black 5-branes each share the same phase structure, one expects that
the charge parameter-space should be symmetric with respect to $q_1$
and $q_5$ for the phase structure. In other words, the dotted
diagonal line of the square connecting the origin and the opposite
corner in Fig. 1 divides the parameter space into two equivalent
regions. This is further support by the critical charges given in
\eqn{criticalc} where $q_{1c}$ and $q_{5c}$ do appear symmetric and
have a symmetric point at the intersection point $(\sqrt{6}/16,
\sqrt{6}/16)$ between the diagonal line and the critical line (also
at the middle of the critical line). For this reason and later for
comparison with black 5-branes, we will focus, from now on, on the
lower half of the region including the diagonal line, i.e. the
region enclosed by the diagonal line, the $q_5$-axis and the $q_5 =
1$ line plus the diagonal line itself. For the critical line, we
will choose in \eqn{criticalc} the plus sign for $q_{5c}$ and minus
sign for $q_{1c}$ with $2/3 \le f_c < 1$ (Note that the symmetric
critical point occurs at $f_c = 2/3$). For convenience, we denote
the region having the phase structure similar to the $q > q_c$ case
for black p-branes  when $\tilde d > 2$ as ``upper region" and the
one containing the van der Waals-Maxwell liquid-gas type as ``lower
region". We now come back to give more evidence in support of our
above claims about the phase structure in ``upper region" and in
``lower region" of the charge space. We cannot give a complete
analysis for the present case, in a spirit similar to that in
\cite{Lu:2010xt}, due to the complexity of the first equation in
\eqn{criticalcd}. However, we know here already the critical line
and so either ``upper region" or ``lower region" must correspond to
one or the other, but not both, phase structure mentioned above. So
selecting a few sampling points in each region will be sufficient to
confirm our above claims. For this, we consider three points $(q_1,
q_5) = (0.088,0.765), (0.153, 0.765), (0.442, 0.765)$ in ``upper
region" and one critical point $ (q_1, q_5) = (0.153, 0.153)$, the
symmetric point on the critical line. As shown in Fig. 2,  there
doesn't exist any extremum of $b_{q_1, q_5} (f)$ for any of three
points selected and for each given $\bar b$, there will be a unique
solution of $\bar b = b_{q_1, q_5} (\bar f)$.
\begin{figure}
\psfrag{a}{$f$} \psfrag{b}{$ 4 b_{q_1, q_5} (f)$}
\begin{center}
\includegraphics{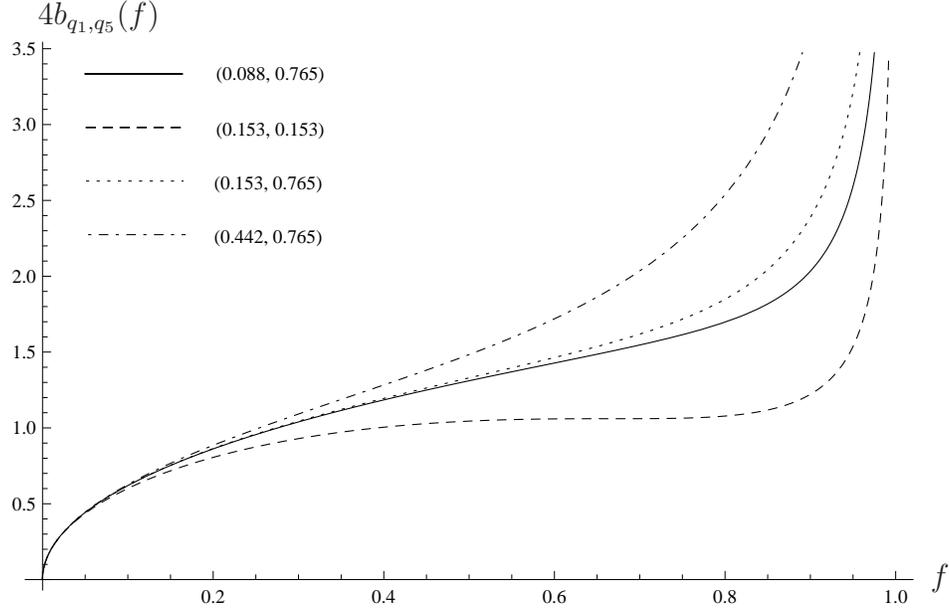}
\end{center}
  \caption{The expected behavior of $b_{q_1, q_5} (f) $ vs $f$ for three selected pairs of ($q_1, q_5$) values in the
  region enclosed by the upper part of diagonal line and the
  lower half of critical line and the $q_5 = 1$ line as shown in Fig. 1. Note that
the pair of $(0.153, 0.153)$ corresponds to the symmetric point on
the critical line.}
\end{figure}
Moreover, given the behavior of $b_{q_1, q_5} (f)$ for $f
\rightarrow 0$ and $\rightarrow 1$ in \eqn{bbvsf}, $ b_{q_1, q_5}
(\bar f)$ increases monotonically for $0 < f < 1$ and its slope is
hence positive. So the corresponding free energy is a local minimum,
therefore stable. The symmetric critical point $(0.153, 0.153)$ in
Fig. 2 does appear to be flat around the point $f = 2/3$ which gives
the critical point. Let us also examine three sampling points $(q_1,
q_5) = (0.031, 0.063), (0.044, 0.088), (0.054, 0.099)$ but now in
``lower region". As shown in Fig. 3, each gives rise to a maximum
and a minimum of $b_{q_1, q_5} (f)$ for $0 < f < 1$, as expected.
When $\bar b$ falls between the minimum and maximum, the equation of
state $\bar b = b_{q_1, q_5} (f)$ will have three solutions $f_1 <
f_2 < f_3$ but only at the smallest $f_1$ or at the largest $f_3$,
the slope of $b_{q_1, q_5} (f)$ is positive and the corresponding
free energy has a local minimum, therefore locally stable. While at
the middle $f_2$, the slope of this function is negative and the
free energy has a maximum there, therefore unstable. For a given
pair of charge $(q_1, q_5)$, when we change $\bar b$ between the
maximum and the minimum, each of these $\bar b$ will give two
locally stable phases but only the one with the lowest free energy
is more stable and the other one is merely meta-stable.
\begin{figure}
\psfrag{a}{$f$} \psfrag{b}{$ 4 b_{q_1, q_5} (f)$}
\begin{center}
\includegraphics{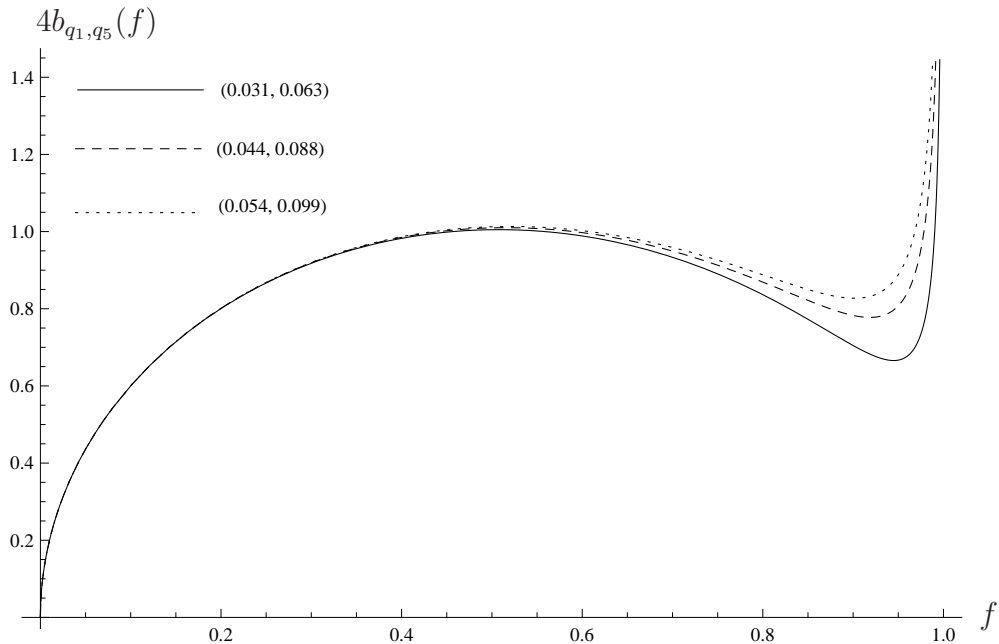}
\end{center}
  \caption{The expected behavior of $b_{q_1, q_5} (f) $ vs $f$ for three selected pairs of ($q_1, q_5$) values in the region enclosed by the
  lower part  of diagonal line and the lower half of critical line and the
$q_5$-axis as shown in Fig. 1.}
\end{figure}
Among these $\bar b$, there exists a special $\bar b_t$ which is
completely determined by these two
 charges and the two locally stable phases have the same free
 energy, therefore the two can co-exist and there can be a phase
 transition between the two. Since $f$ is related to the horizon size
 $x$ via \eqn{variablet} and a larger $f$ corresponds to a smaller $x$, so
 $f_1$ corresponds to a larger horizon size while $f_3$ corresponds to a smaller horizon size.
 Therefore such a phase transition
 involves a change of horizon size, implying a change of entropy of
 the underlying system. So we have this phase transition a
 first-order one. The $\bar b_t$ is just the inverse of the
 corresponding phase transition reduced temperature. Following
 \cite{Lu:2010xt}, when the temperature is lower than the transition
 one, i.e., $\bar b > \bar b_t$, the large $f_3$ (therefore corresponding to a small horzion size)
  phase will have a
 lower free energy, therefore more stable. Otherwise, the small
 $f_1$ phase will have a lower free energy, therefore more stable.
So we also have here a van der Waals-Maxwell liquid-gas phase
structure as anticipated. Here the discussion is similar to the $q <
q_c$ case of black p-branes when $\tilde d > 2$ given in
\cite{Lu:2010xt}. In drawing analogy to the liquid-gas phase
diagram, here the small stable D1/D5 system corresponds to the
liquid phase while the large one to the gas phase.

Given what has been said, an analysis for a special case of $q_1 =
q_5$, in a spirit similar to that in \cite{Lu:2010xt}, is still
possible for the phase structure. This will lend further support to
our above claims. In other words, we now consider the diagonal line
of the charge square in Fig. 1. From the first equation in
\eqn{criticalcd}
 or \eqn{cc}, we have the following equation, which determines the position of the extremality of $b_{q, q} (f)$, as
\be \label{exeqn} (6 - 4 q^2) f^3 - (15 - 16 q^2) f^2 + (12 - 16
q^2) f - 3 = 0, \ee where for simplicity we have set $q_1 = q_5 =
q$. This is a cubic equation and has three roots. We will discuss
the properties of these three roots, following the analysis of
\cite{Lu:2010xt}, to determine the behavior of $b_{q, q} (f)$,
therefore the phase structure.  For this, we have the discriminant
of the above cubic equation as \be \Delta (q) = 144 q^2 (q - 1) (q +
1) (128 q^2 - 3).\ee Given $0 < q < 1$, we know that the
discriminant can only vanish if $q = q_c = \sqrt{6}/16$, which is
nothing but the expected symmetric (middle) point on the critical
line for which the two positive roots coincide at $f_c = 2/3$. The
third root is actually negative, therefore not in the region of our
interest, since the sign of the product of three roots is given by
that of the ratio of the coefficient of $f^3$ to the constant term
in \eqn{exeqn}, which is negative.

When $1 > q > q_c$, we have $\Delta (q) < 0$ and so we have now a
pair of complex conjugate roots and one negative root. None of them
are in the region of our interest. In other words, $b_{q, q} (f)$
doesn't have any extremality in the region $0 < f < 1$ and so $b_{q,
q} (f)$ must increase monotonically in this region given its
behavior for $f\rightarrow 0$ and $f\rightarrow 1$ in \eqn{bbvsf}.
This further implies its slope being positive and a unique solution
from the equation of state $\bar b = b_{q, q} (\bar f)$. So the free
energy at each equilibrium point $f = \bar f$ has a minimum,
therefore stable. This is nothing but the behavior of the $q
> q_c$ case of black p-branes when $\tilde d
> 2$ considered in \cite{Lu:2010xt}.

When $q < q_c$, we have $\Delta (q) > 0$ and so the three roots are
all real.  Note that the left side of equation \eqn{exeqn} takes a
value of $- 3$ at $f = 0$ and a value of $- 4 q^2$ at $f = 1$, both
being negative at the two ends of $0 < f < 1$, therefore the number
of roots, if they exist at all in the region of $0 < f < 1$, must be
even (two or none). Given the fact about the existence of critical
point $f_c = 2/3$, which results from the coincidence of two
positive roots when $q \rightarrow q_c$, this implies therefore the
existence of two positive roots indeed in the region of $0 < f < 1$.
Here the analysis goes parallel to that given in \cite{Lu:2010xt}
for $q < q_c$ case of black p-branes when $\tilde d
> 2$ and so we expect that the underlying phase structure contains also the van der
Waals-Maxwell liquid-gas type.

As in \cite{Lu:2010xt} for black p-branes with $\tilde d > 2$, the
extremal case (corresponding to taking $f = 1$) with the same
boundary data has always larger free energy than the non-extremal
case, therefore can never be a stable thermodynamical state in
canonical ensemble\footnote{With the same boundary data, the
extremal case has its free energy $\bar F_{\rm extremal} = \bar
I^{\rm extremal}_E /\bar b = 2 \,(q_1 + q_5)$ following the
discussion given in \cite{Lu:2011da} while the non-extremal one has
its on-shell free energy $\bar F_{\rm non-extremal} (\bar f) = - 2\,
\bar f^{-1/2}[ \bar f - 3 \bar f^{1/2} + 2 - (\sqrt{(1 - \bar f)^2 +
4 q_1^2 \bar f} + \sqrt{(1 - \bar f)^2 + 4 q_5^2 \bar f} )/2]$ where
we have used the equation of state \eqn{eosf} and $\bar f$ is a
solution from \eqn{eosf} for each given $\bar b$. We expect in
general $0 < \bar f < 1$ but not every its value, when in ``lower
region", gives a stable phase. Note that $\bar F_{\rm non-extremal}
(\bar f = 0) = - \infty$ and $\bar F_{\rm non-extremal} (\bar f = 1)
= 2 (q_1 + q_5)$ at its two limiting ends. We would like to point
out that though both $\bar F_{\rm non-extremal} (\bar f = 1)$ and
$\bar F_{\rm extremal}$ have the same value $2\, (q_1 + q_5)$ but
the extremal one has this value, independent of $\bar f$ or $\bar
b$, following \cite{Gibbons:1994ff, Hawking:1994ii, Hawking:1995fd,
Teitelboim:1994az}. One can check directly for the on-shell free
energy $d \bar F_{\rm non-extremal}/d \bar f \sim d b_{q_1, q_5}
(\bar f) /d \bar f$. Note that $\bar F_{\rm non-extremal} (\bar f)$
has its value as a local minimum at $f = \bar f$ and therefore $d
b_{q_1, q_5} (\bar f) /d \bar f > 0$. This implies that the on-shell
free energy has its $d \bar F_{\rm non-extremal}/d \bar f > 0$ for
locally stable phase.  In other words, except for at the phase
transition point if relevant at all, the on-shell free energy for
stable phase is a monotonically increasing function of variable
$\bar f$ and so this further implies that the end value of $\bar
F_{\rm non-extremal} (\bar f = 1) = 2 (q_1 + q_5)$ is the largest
one. Therefore the extremal one has always the largest free energy
given what has been said above. The similar line of argument was
also employed in \cite{Lu:2011da} for the case of black p-branes
with $\tilde d
> 2$.  Note that for the present extremal case, the total free energy is simply
the sum of the free energies due to the delocalized strings and
D5-branes while this is not true for the non-extremal case. This
lowered free energy for non-extremal case should be due to the
attractive interaction between the delocalized strings and D5-branes
while for extremal case, the delocalized strings don't interact with
the D5-branes since their brane dimensions differ by four. We will
elaborate this point later in this section.}.

In summary, with the addition of  delocalized charged D-strings, the
phase structure of charged black D5-branes with $\tilde d = 2$ as
described in \cite{Lu:2010xt} has been qualitatively changed to the
one similar to that of black p-branes with $\tilde d > 2$, a van der
Waals-Maxwell liquid-gas type, even though the delocalized charged
 black D-strings alone share the same phase structure as the original
charged black D5-branes, without such a type phase structure. The
present phase structure is actually even more richer since we have
now a critical line instead of a critical point which separates a
two-dimensional charge square into two regions with each giving rise
to a different phase behavior as described earlier. Each point on
the critical line represents a second-order phase transition. Any
continuous curve crossing the critical line in the two-dimensional
charge square gives rise to a phase structure like that of black
p-branes with $\tilde d > 2$, having a first-order phase transition
line which terminates at a second-order phase transition point,
corresponding to the crossing point between this charge curve and
the critical line.

At one hand, the delocalized charged black D-strings share the same
phase structure as the charged black D5-branes \cite{Lu:2010xt} with
$\tilde d = 2$. This seemingly indicates that the delocalized
charged black D-strings behave, at least in terms of thermodynamics,
like the charged black D5-branes. Therefore one would naively expect
that when the two are combined to form black D1/D5 system, its phase
behavior remains the same as before. But on the other hand, to our
surprise, the resulting phase structure is actually completely
different as revealed in this study. As hinted already in footnote
12 given previously, this is due to the attractive interaction
between the black D-strings and the black D5-branes, which is quite
different from that between black D5-branes. We will elaborate this
further in what follows.

When both D1 and D5 are extremal or BPS, it is well-known that there
exist no interaction between any two of these branes, whether they
are both D1 or both D5 or one D1 and the other D5. As given in
footnote 12, the total free-energy of extremal D1/D5 system is just
the sum of constituent brane contributions and the reason behind is
just the mentioned no interactions among these branes. For charged
black D-strings or/and D5-branes, the story is different. For either
type of branes, the net interaction is attractive since the
attractive gravitational one due to brane masses overtakes the
repulsive one due to their R-R charges. For each given type of
branes, the corresponding black configuration remains the same in
form and so the form of its Euclidean action remains also the same,
independent of its charge. This can be easily seen, for example,
from \eqn{ractionf} when we set either $q_1$ or $q_5$ to zero. When
both $q_1$ and $q_5$ are non-zero, the resulting form is completely
different from that when only one of charges is non-zero. This is
due to the following reasons. D-string R-R charges don't interact
with D5 ones and the interaction between them are due to their
masses. This attractive interaction is different from that between
D-strings or between D5-branes and it is not additive.

The above understanding is based on the possible interactions which
appear in the system. In terms of thermodynamics, the understanding
is a bit different and the non-additive contributions appear in both
the internal energy and the entropy of the underlying system which
can be seen from, for example, the reduced action \eqn{ractionf}.
Let us examine this action in a bit detail. First, if we set either
$q_1$ or $q_5$ zero in the action, we end up with the same formula
since the action is symmetric with respect to $q_1$ and $q_5$ and
this confirms that the
 delocalized charged black D-strings share the same phase structure
as the charged black D5-branes. Second, when both $q_1$ and $q_5$
are non-zero, the reduced action is not the sum of the action due to
D-strings and that due to D5-branes.  This occurs not only in the
first term (the internal energy multiplied by the common $\bar b$
factor) but also in the second term (entropy) of the reduced action
\eqn{ractionf}. However, for the corresponding extremal case (taking
$f = 1$ in the action), the reduced action is indeed the sum of the
above two corresponding extremal actions, as already indicated in
footnote 12.

In other words, the dramatic change of the phase structure for
either charged black D5-branes or the  delocalized charged black
D-strings, when the other type of branes are added, are caused by
the interaction between these two types of branes in the D1/D5
system considered. In comparison with the phase structures of black
p-branes considered in \cite{Lu:2010xt}, this change represents a
phase structure change in characterization from the $\tilde d = 2$
case to $\tilde d > 2$ case. To be more precise, such a change is
referred to any given continuous charge curve in the lower half of
the charge square (including the diagonal line) in Fig. 1, starting
on the $q_5$-axis where $q_1 = 0$ (corresponding to $\tilde d = 2$),
then in ``lower region", crossing the critical line and finally in
``upper region" (when $q_1 \neq 0$, this curve appears to give rise
to an effective $\tilde d > 2$).

The above discussion appears to imply that adding the delocalized
charged D-strings to charged black D5-branes changes effectively its
transverse dimensions $\tilde d$ from $\tilde d = 2$ to a $\tilde d
> 2$. This can be a possibility, given that the characteristic
phase structure and critical phenomena for the charged black
p-branes depends only on the dimensionality of transverse dimensions
$\tilde d$. To confirm the possibility would require that the
present D1/D5 system can be modeled effectively by a charged black
brane system with an effective $\tilde d > 2$ but this will be
beyond the scope of this paper. Instead, we will give a limited
consideration by providing one piece of evidence in support of this
connection.

At least for the charged black p-branes, the underlying phase
structure is closely related to the corresponding critical point and
the latter is completely determined by the transverse dimensions
$\tilde d$, not the details of the underlying system. We also know
that the charged black p-branes have a completely different phase
structure for $\tilde d > 2$ and $\tilde d = 2$, and the
corresponding critical parameters such as $q_c$ and $b_c$ show a
pattern of decrease when $\tilde d$ increases from $\tilde d = 2$ to
$\tilde d = 7$ \cite{Lu:2010xt} as shown in the following table:
\begin{center}
\begin{tabular}{|c|c|c|}
\hline
$\tilde d$ & $q_c$ & $b_c$\\
\hline\hline
2& 0.333333&0.288675\\
\hline
3 & 0.141626 &  0.199253\\
\hline
4 & 0.090672 & 0.159921\\
\hline
5 & 0.064944 & 0.134632\\
\hline
6 & 0.049599 & 0.116698\\
\hline
7 & 0.039529 &  0.103210\\
\hline
\end{tabular}
\centerline{}
 \centerline{Table 1. The critical parameters $q_c$ and
$b_c$ for black p-branes with $\tilde d \ge 2$}
\end{center}
In other words, the decrease of both $q_c$ and $b_c$ corresponds to
the increase of $\tilde d$. For the present D1/D5 system, when $q_1
= 0$, i.e., without the presence of the delocalized D-strings, this
system is nothing but the above charged black 5-branes with $\tilde
d = 2$. We now deform this charged black 5-branes by adding
delocalized D-strings and we end up with a richer phase structure,
as discussed earlier, with a critical line described by a pair of
critical charge $(q_{1c}, q_{5c})$. The present critical
parameters\footnote{As stated earlier, we have chosen the lower half
of the critical line in Fig. 1 for the sake of comparing with the
black D5-branes. Note that $q_{1c} = \sqrt{6}/16$ corresponds to the
symmetric point.} $q_{1c}, q_{5c}, b_c$ for $2/3 \le f_c \le 1$ are
given in Table 2.
\begin{center}
\begin{tabular}{|c|c|c|c|}
\hline
$f_c$ & $q_{1c}$ &$q_{5c}$ & $b_c$\\
\hline\hline
1& 0 &0.333333&0.288675\\
\hline
0.96 & 0.000239 & 0.332405 &0.288265\\
 \hline
0.92 & 0.001431 & 0.329431 &0.287193\\
\hline
0.88&0.004216&0.324028&0.286295\\

\hline
0.84&0.009350&0.315613&0.282938\\

\hline
0.80&0.017906&0.303256&0.279845\\

\hline
0.76&0.031669&0.285292&0.276135\\

\hline
0.72&0.054382&0.258084&0.271828\\

\hline
2/3&0.153093&0.153093&0.265165\\
 \hline
\end{tabular}
\centerline{} \centerline{Table 2. The present critical parameters
$q_{1c}, q_{5c}$ and $b_c$ for $ 2/3 \le  f_c \le 1$.}
\end{center}
 Note that $f_c = 1$ corresponds to the simple black
D5-branes with $\tilde d = 2$ without the delocalized D1, i.e.
$q_{1c} = 0$. The corresponding $q_{5c}$ and $b_c$ values are
precise the ones given in the Table 1. Examining the pattern of the
critical parameters, we see first that the critical $b_c$ decreases
in the same direction as $f_c$ does. If such an effective
description is assumed, the correlation between $b_c$ and $\tilde d$
given in Table 1 mentioned above implies an increase of the present
effective $\tilde d$,  from the original $\tilde d = 2$ to a new
$\tilde d > 2$ for $f_c < 1$. One expects that the largest effective
$\tilde d$ corresponds to $f_c = 2/3$ and $b_c = 0.265165$ given in
Table 2. As for $ f_c < 1$, we have a non-vanishing $q_{1c}$ and as
we see from Table 2, the present $q_{5c}$ value is no longer its
original one $q_{5c} = 1/3$ (i.e., corresponding to $q_{1c} = 0$).
This further implies that the new $q_{5c}$ already takes the effect
of the non-vanishing $q_{1c}$ into consideration. So this new
 $q_{5c}$ qualitatively appears as an effective $q_c$ given in Table 1.
 Its decreasing pattern correlates with that of $b_c$ lends further
 support to this assertion.

The largest $q_{1c}$ corresponds to the symmetric point in Figure 1
where $q_{1c} = q_{5c} = 0.153093$ (or the exact value of
$\sqrt{6}/16$) and $f_c = 2/3$. Here we have $b_c = 0.265165$ (or
the exact value of $3 \sqrt{2}/16$), still far away from $b_c =
0.199253$ of $\tilde d = 3$ given in Table 1. So we expect that the
effective $\tilde d > 2$ but still far away from $\tilde d = 3$, if
such an effective description exists indeed. So a brane charge has a
potential in generating effectively a fraction of spatial dimension.

\section{Discussion and conclusion}
In this paper, we explore means which can be used potentially to
change qualitatively the phase structure of black p-branes in
canonical ensemble. Following \cite{Lu:2010xt}, we know that the
charged black branes with their $\tilde d = 2$ serve as a borderline
between the black branes with $\tilde d = 1$ and those with $\tilde
d > 2$ in phase structure. So it should be much easier to realize
such a change if we focus on the $\tilde d = 2$ branes, which in $D
= 10$ are black D5 branes (or NS5-branes). The simplest system to
consider is the black D1/D5. The  delocalized charged black
D-strings alone are found to have the exact same phase structure as
the charged black D5-branes without a van-der Waals-Maxwell
liquid-gas type structure. Given what we know about D1 and D5 in the
BPS D1/D5 case, one would naively think that when the two are
combined, the resulting phase structure should remain as before.
However, it turns out that the resulting one is much richer and
contains the liquid-gas type, with now a critical line instead of a
critical point. We find that the physical reason for this to occur
is the existence of a non-vanishing interaction between charged
black D-strings and charged black D5-branes, which differs from that
between the delocalized black D-strings or between black D5-branes.

The reduced D1 charge $q_1$ and the reduced D5 brane charge $q_5$
form the interior of a two-dimensional charge square. Given that the
 delocalized charged black D-strings and the charged black D5-branes
each share the same phase structure, therefore so long the phase
structure is concerned, we expect that there is a symmetry between
$q_1$ and $q_5$, as demonstrated in this paper. Therefore the
fundamental region in the charge space is either of the two
equivalent sub-regions so obtained by this symmetry, i.e., the
region defined on either side of the diagonal line connecting the
origin and its opposite corner of the square, plus the diagonal line
excluding its two end points. In the paper, we choose for
convenience the one nearby  the $q_5$-axis. There exists a critical
line in this finite two-dimensional charge space with its one end
approaching $q_1$-axis and the other approaching $q_5$-axis but
neither of them reaching the corresponding axis. If we focus on the
fundamental region, the corresponding critical line is the one with
its one end terminating on the diagonal line and the other end
approaching the $q_5$-axis but not reaching it. Then this critical
line separates the fundamental region into two parts. In the
so-called ``upper region" (defined earlier), any point in the charge
space defines a unique thermodynamical stable black D1/D5 for given
boundary data, similar to the $q > q_c$ case for the black p-branes
with their $\tilde d > 2$. In the ``lower region", any point in the
charge space defines a van der Waals-Maxwell liquid-gas type phase
diagram for the black D1/D5 system for which there usually exist two
local thermodynamical stable D1/D5 states, one small and one large
in horizon size, similar to the $q < q_c$ case for black p-branes
with also their $\tilde d > 2$. The small one is analogous to the
liquid phase while the large one to gas phase. Which one is more
stable depends on the given boundary data and the lower cavity
temperature favors the small one while higher temperature favors the
large one when compared with the so-called transition temperature.
For each given point in this region, there is a unique transition
temperature determined by this point and at which the large and
small black D1/D5 systems have the same free energy, therefore can
co-exist and there is a first-order phase transition since any
exchange between these two states involves a change of entropy.

For any charge curve in ``lower region" terminating on the critical
line, we will have a first-order phase transition line ending on a
second-order phase transition point when the charge point $(q_1,
q_5)$ reaches its critical point $(q_{1c}, q_{5c})$ on the critical
line. So any charge curve crossing the critical line in the
fundamental region defined above will have a phase structure like
that of black p-branes with their $\tilde d > 2$. This naturally
leads to a speculation that the presence of $q_1$ can have a
consequence, at least effectively, of increasing the original
D5-brane $\tilde d = 2$ to a $\tilde d > 2$ slightly.  So to certain
extent, we see that a brane charge can effectively give rise to a
fraction of a spatial dimension.

\section*{Acknowledgements:}

We would like to thank the anonymous referee for fruitful
suggestions which help us to improve the manuscript. We acknowledge
support by grants from the Chinese Academy of Sciences and grants
from the NSF of China with Grant No : 10975129 and 11235010.

\section*{Appendix}
We in this appendix consider the general charged black D(p - 4)/Dp
system for $4 \le p \le 6$. The corresponding black configuration in
Euclidean signature is \bea \label{gbc} ds^{2} &=&{G_{p-4}}^{\frac{p
- 11}{8}}{G_{p}}^{\frac{p - 7}{8}}\left(f dt^{2} + \sum_{i = 1}^{p -
4} d x_i^2\right) + {G_{p - 4}}^{\frac{p - 3}{8}}{G_{p}}^{\frac{p -
7}{8}}\sum_{i= p - 3}^{p} dx_i^{2} \nn &\,& + {G_{p - 4}}^{\frac{p -
3}{8}}{G_{p}}^{\frac{p + 1}{8}} \left(\frac{d\rho^{2}}{f} + \rho^{2}
d\Omega_{8 - p}^{2} \right)\nn A_{[p - 3]}&=&-ie^{\alpha(p - 4)
\phi_{0}/2}\left[\tanh\theta_{p - 4}-(1- G_{p -
4}^{-1})\coth\theta_{p - 4}\right]dt\wedge dx^1\wedge \cdots \wedge
d x^{p - 4}\nn A_{[p + 1]}&=&-ie^{\alpha(p)
\phi_{0}/2}\left[\tanh\theta_{p}-(1-
{G_{p}}^{-1})\coth\theta_{p}\right]dt\wedge dx^{1}...\wedge
dx^{p}\nn e^{2(\phi-\phi_{0})}&=&G_{p - 4}^{- \alpha(p - 4)}
G_{p}^{-\alpha (p)}, \eea where the metric is in Einstein frame,
each form field is obtained following \cite{Braden:1990hw,
Lu:2010xt} in such a way that the form field vanishes at the horizon
so that it is well defined in the local inertial frame, $\phi_0$ is
the asymptotic value for the dilaton and \be {G}_{p -
4,p}=1+\frac{\rho_{0}^{7 - p}\sinh^{2}\theta_{p - 4, p}}{\rho^{7 -
p}}, \qquad f=1-\frac{\rho_{0}^{7 - p}}{\rho^{7 - p}}.\ee The
horizon is at $\rho = \rho_0$ with a curvature singularity behind at
$\rho = 0$. In the above, \be \alpha (p) = \left\{\begin{array}{cc}
\frac{p - 3}{2} &{\rm for\,\, Dp-brane},\\
- \frac{p - 3}{2} &{\rm for \,\, NSp-brane}.\\
\end{array}\right. \ee
Once again, for convenience, we generalize the radial coordinate
transformation \eqn{vd} from $\rho$ to $r$  to the present case as
\be \rho^{7 - p} = r^{7 - p} - r_-^{7 - p}, \qquad \rho_0^{7 - p} =
r_+^{7 - p} - r_-^{7 - p},\ee where $r_+^{7 - p} \equiv  \rho_0^{7 -
p} \cosh^2 \theta_p \ge r_-^{7 - p} \equiv \rho_0^{7 - p}
\sinh^2\theta_p$. In terms of the new radial coordinate $r$, the
horizon is now at $r = r_+$ and the singularity is at $r = r_-$. The
configuration \eqn{gbc} can now be expressed as \bea \label{gbc-r}
ds^{2} &=&{G_{p-4}}^{\frac{p - 11}{8}}{\triangle_-}^{\frac{7 -
p}{8}}\left(\frac{\triangle_+}{\triangle_-} dt^{2} + \sum_{i = 1}^{p
- 4} d x_i^2\right) + {G_{p - 4}}^{\frac{p -
3}{8}}{\triangle_-}^{\frac{7 - p}{8}}\sum_{i= p - 3}^{p} dx_i^{2}
\nn &\,& + {G_{p - 4}}^{\frac{p - 3}{8}}{\triangle_-}^{- \frac{p +
1}{8}} \left(\frac{\triangle_-^{\frac{p - 5}{7 -
p}}}{\triangle_+}\,d r^2 + r^{2} \triangle_-^{\frac{2}{7 - p}}
d\Omega_{8 - p}^{2} \right)\nn A_{[p - 3]}&=&-ie^{\alpha(p - 4)
\phi_{0}/2}\left[\tanh\theta_{p - 4}-(1- G_{p -
4}^{-1})\coth\theta_{p - 4}\right]dt\wedge dx^1\wedge \cdots \wedge
d x^{p - 4}\nn A_{[p + 1]}&=&-ie^{\alpha(p)
\phi_{0}/2}\left[\tanh\theta_{p}-(1-
\triangle_-)\coth\theta_{p}\right]dt\wedge dx^{1}...\wedge dx^{p}\nn
e^{2(\phi-\phi_{0})}&=&G_{p - 4}^{- \alpha(p - 4)}
\triangle_-^{\alpha (p)}, \eea where now \be G_{p - 4} = 1 + \left(1
- \frac{\triangle_+}{\triangle_-}\right) \sinh^2\theta_{p - 4},
\qquad G_p = \triangle_-^{-1}.\ee Here \be \triangle_\pm = 1 -
\frac{r_\pm^{7 - p}}{r^{7 - p}}.\ee In the above metric, the
Euclidean time is periodic with a periodicity \be \label{gbeta}
\beta^* = \frac{4 \pi r_+ \cosh\theta_{p - 4}}{7 - p} \left[1 -
\left(\frac{r_-}{r_+}\right)^{7 - p}\right]^{- \frac{5 - p}{2(7 -
p)}},\ee which is the inverse of temperature of the system at $r =
\infty$. By the same token, we place this system in a cavity at $r =
r_B$. As before, $r_B$ is merely a coordinate radius, not the
physical one which, from the metric, is defined as  \be \bar r_B =
r_B G_{p - 4}^{\frac{p - 3}{16}} \triangle_-^{\frac{(p - 3)^2}{16 (7
- p)}}.\ee The physical parameters \be \bar r_\pm = r_\pm G_{p -
4}^{\frac{p - 3}{16}} \triangle_-^{\frac{(p - 3)^2}{16 (7 - p)}}.\ee
In terms of these physical quantities,  $\triangle_\pm$ once again
keep their respective forms unchanged \be \triangle_\pm = 1 -
\frac{r_\pm^{7 - p}}{r^{7 - p}} = 1 - \frac{\bar r_\pm^{7 - p}}{\bar
r^{7 - p}}, \ee and the inverse of local temperature at $\bar r_B$
is \bea \label{ilt} \beta (\bar r_B) &\equiv & \triangle_+^{1/2}
G_{p - 4}^{\frac{p - 11}{16}} \triangle_-^{- \frac{p + 1}{16}}
\beta^*,\nn & = & \frac{4\pi \bar r_+}{7 - p}
\left(\frac{\triangle_+}{\triangle_-}\right)^{1/2}\left(\frac{\bar
r_+}{\bar r_B}\right)^{\frac{5 - p}{2}} \left(1 -
\frac{\triangle_+}{\triangle_-}\right)^{\frac{p - 5}{2(7 - p)}}
\left(1 + \frac{1 - G_{p - 4}^{ -
1}}{\frac{\triangle_-}{\triangle_+} - 1}\right)^{1/2},\eea where we
have used the expression for $\beta^*$ given in \eqn{gbeta}. The
respective charge can be computed for this configuration, following
\be \label{gc} Q_p = \frac{i}{\sqrt{2} \kappa} \int_{S_\infty^{8 -
p}} e^{- \alpha (p) \phi} \ast F_{p + 2},\ee as \bea \label{gc} Q_{p
- 4} &=& \frac{(7 - p) \bar V_4 \Omega_{8 - p}}{\sqrt{2} \kappa}
e^{- \alpha (p - 4) \bar \phi/2} \bar r_B^{7 - p} \left(1 - G_{p -
4}^{- 1}\right)^{1/2} \left(1 - \frac{\triangle_+}{\triangle_-} G_{p
- 4}^{ - 1}\right)^{1/2},\nn Q_p &=& \frac{(7 - p) \Omega_{8 -
p}}{\sqrt{2}\kappa} e^{- \alpha (p) \bar \phi/2} \left(\bar r_+ \bar
r_-\right)^{\frac{7 - p}{2}},\eea where we have expressed relevant
quantities in terms of either the corresponding physical ones or the
fixed values on the cavity. Note that all the above formulas, when
set $p = 5$, reduce to those of the D1/D5 system as expected.

Following what we did for the D1/D5 system, we have the Euclidean
action as \bea \label{gaction}I_E &=& \beta E - S\nn &=& -
\frac{\beta \bar V_p \Omega_{8 - p}}{\kappa^2} \bar r_B^{7 - p}
\left[\frac{7 - p}{2} \sqrt{\triangle_+ \triangle_-} + \frac{9 -
p}{2}\sqrt{\frac{\triangle_+}{\triangle_-}} - \frac{7 - p}{2}
\sqrt{\frac{\triangle_+}{\triangle_-}} \left(1 - G_{p - 4}^{-
1}\right) - (8 - p)\right]\nn &\,& - \frac{2 \pi \bar r_+ \bar V_p
\Omega_{8 - p}}{\kappa^2} \bar r_B^{7 - p} \left(\frac{\bar
r_+}{\bar r_B}\right)^{\frac{5 - p}{2}} \left(1 -
\frac{\triangle_+}{\triangle_-}\right)^{\frac{9 - p}{2(7 - p)}}
\left(1 + \frac{1 - G_{p - 4}^{- 1}}{\frac{\triangle_-}{\triangle_+}
- 1} \right)^{\frac{1}{2}},\eea which reduces to the D1/D5 one
\eqn{actiond1d5} when $p = 5$. From this action, one can read the
corresponding internal energy $E$ and entropy $S$, respectively.
Similarly, we define the reduced charge\bea\label{gcc} \tilde Q_{p -
4} &\equiv& \frac{\sqrt{2} \kappa Q_{p - 4} e^{\alpha(p - 4) \bar
\phi/2}}{(7 - p) \bar V_4 \Omega_{8 - p} \bar r^{7 - p}_B}\nn & = &
\left(1 - G_{p - 4}^{- 1}\right)^{1/2} \left(1 -
\frac{\triangle_+}{\triangle_-} G_{p - 4}^{-1}\right)^{1/2} < 1,\nn
\tilde Q_p &\equiv& \left(\frac{\sqrt{2} \kappa Q_p e^{\alpha (p)
\bar\phi/2}}{(7 - p) \Omega_{8 - p}}\right)^{\frac{1}{7 - p}} =
\sqrt{\bar r_+ \bar r_-},\eea where we have used \eqn{gc} for $Q_{p
- 4}$ and $Q_p$, respectively. From the second expression of above,
we can express $\bar r_-$ in terms of $\bar r_+$ and further we can
solve $G_{p - 4}$ in terms of $\bar r_+$ from the first equation. We
end up with \bea \label{cs-solution} \bar r_- &=& \frac{\tilde
Q_p^2}{\bar r_+}, \nn 1 - G_{p - 4}^{- 1} &=&
\frac{1}{2}\left[\sqrt{\left(\frac{\triangle_-}{\triangle_+} -
1\right)^2 + 4 \tilde Q_{p - 4}^2 \frac{\triangle_-}{\triangle_+}} -
\left(\frac{\triangle_-}{\triangle_+} - 1\right)\right].\eea Since
in canonical ensemble, $\beta, \bar r_B, \bar V_p, \tilde Q_{p - 4}$
and $\tilde Q_p$ are all fixed,  the only variable in the action
\eqn{gaction} is now the physical horizon size $\bar r_+$. For
simplicity, just like before, we define the following reduced action
\bea \label{graction} \bar I_E &\equiv& \frac{\kappa^2 I_E}{2\pi
\bar r_B^{8 - p} \bar V_p \Omega_{8 - p}}\nn &=&- \frac{\beta}{2\pi
\bar r_B}\left[\frac{7 - p}{2} \sqrt{\triangle_+ \triangle_-} +
\frac{9 - p}{2}\sqrt{\frac{\triangle_+}{\triangle_-}} - \frac{7 -
p}{2} \sqrt{\frac{\triangle_+}{\triangle_-}} \left(1 - G_{p - 4}^{-
1}\right) - (8 - p)\right]\nn &\,& -  \left(\frac{\bar r_+}{\bar
r_B}\right)^{\frac{7 - p}{2}} \left(1 -
\frac{\triangle_+}{\triangle_-}\right)^{\frac{9 - p}{2(7 - p)}}
\left(1 + \frac{1 - G_{p - 4}^{- 1}}{\frac{\triangle_-}{\triangle_+}
- 1} \right)^{\frac{1}{2}},\eea where we have used the action
\eqn{gaction} and which reduces to the D1/D5 one \eqn{reducedaction}
when $p = 5$. Once again for simplicity, we define \be x =
\left(\frac{\bar r_+}{\bar r_B}\right)^{7 - p} < 1, \quad \bar b =
\frac{\beta}{4\pi \bar r_B}, \quad q_p = \left(\frac{\tilde
Q_p}{\bar r_B}\right)^{7 - p} < x < 1, \quad q_{p - 4} = \tilde Q_{p
- 4} < 1.\ee In terms of these reduced quantities, we have
\be\label{dpdm}  \triangle_+ = 1 - \left(\frac{\bar r_+}{\bar
r_B}\right)^{7 - p} = 1 - x, \quad \triangle_- = 1 -
\left(\frac{\bar r_-}{\bar r_B}\right)^{7 - p} = 1 -
\frac{q^2_p}{x},\ee where in the second expression, we have used the
first equation for $\bar r_-$ from \eqn{cs-solution}. Then the
reduced action \eqn{graction} can be expressed as \bea
\label{gractionv} \bar I_E  &=&- 2\, \bar b \left[\frac{7 - p}{2}
\sqrt{\triangle_+ \triangle_-} + \frac{9 -
p}{2}\sqrt{\frac{\triangle_+}{\triangle_-}} - \frac{7 - p}{2}
\sqrt{\frac{\triangle_+}{\triangle_-}} \left(1 - G_{p - 4}^{-
1}\right) - (8 - p)\right]\nn &\,& -  x^{1/2} \left(1 -
\frac{\triangle_+}{\triangle_-}\right)^{\frac{9 - p}{2(7 - p)}}
\left(1 + \frac{1 - G_{p - 4}^{- 1}}{\frac{\triangle_-}{\triangle_+}
- 1} \right)^{\frac{1}{2}},\eea where we have used the second
equation for $1 - G_{p - 4}^{-1}$ in \eqn{cs-solution} and the
expressions in \eqn{dpdm} for $\triangle_\pm$, respectively. From
\be \frac{d\bar I_E}{d x} = 0, \ee we have the equation of state \be
\label{geos}\bar b = b_{q_{p - 4}, q_p} (\bar x),\ee where \be
\label{gbf} b_{q_{p - 4}, q_p} (x) = \frac{x^{1/2}}{7 - p}
\left(\frac{\triangle_+}{\triangle_-}\right)^{1/2} \left(1 -
\frac{\triangle_+}{\triangle_-}\right)^{\frac{p -5}{2(7 -
p)}}\left(1 + \frac{1 - G_{p - 4}^{ -
1}}{\frac{\triangle_-}{\triangle_+} - 1}\right)^{1/2}, \ee i.e., the
inverse of local temperature \eqn{ilt} divided by $4\pi \bar r_B$.
The above equation of state $\bar b = b_{q_{p - 4}, q_p} (\bar x)$
is just the thermal equilibrium condition of the underlying system
with the cavity. As always, the behavior of function $b_{q_{p - 4},
q_p} (x)$ determines the underlying phase structure of the system.
Actually, \be \left.\frac{d^2\bar I_E}{d x^2}\right|_{x = \bar x}
\sim - \frac{d b_{q_{p - 4}, q_p} (\bar x)}{d \bar x}, \ee where
$\bar x$ is a solution of the equation of state given in \eqn{geos}.
So this equation once again says that the underlying system has its
free energy a local minimum at $x = \bar x$ if the slope of $b_{q_{p
- 4}, q_p} (x)$ is negative there, therefore giving its local
stability in thermodynamics.

Let us examine the behavior of $b_{q_{p - 4}, q_p}$ a bit in detail.
As expected, if we set $q_{p - 4} = 0$, it should give the
corresponding one $b_q (x)$ for the black p-branes in $D = 10$
\cite{Lu:2010xt} with $\tilde d = 7 - p$ (here $q = q_p$). One can
check with a bit algebra that this is indeed true, i.e.,
\bea\label{pbrane} b_{0, q_p} (x) &=& \frac{x^{1/2}}{7 -
p}\left(\frac{\triangle_+}{\triangle_-}\right)^{1/2} \left(1 -
\frac{\triangle_+}{\triangle_-}\right)^{\frac{p -5}{2(7 - p)}}\nn
&=& \frac{x^{\frac{1}{7 - p}}}{7 - p} \left(1 - x\right)^{1/2}
\left(1 - \frac{q^2_p}{x^2}\right)^{\frac{p - 5}{2(7 - p)}} \left(1
- \frac{q_p^2}{x}\right)^{-\frac{1}{7 - p}} = b_{q_p} (x),\eea where
we have used the explicit expressions of $\triangle_\pm$ in
\eqn{dpdm} and $1 - G_{p - 4}^{-1} = 0$ when $q_{p - 4} = 0$. Note
that $q_p < x < 1$. Let us examine what happens if we set $q_p = 0$
in $b_{q_{p - 4}, q_p} (x)$. We have now \be  b_{q_{p - 4}, 0} (x) =
\frac{x^{\frac{p - 5}{2(7 - p)}}}{7 - p} \left(1 - x\right)^{1/2}
\left(\frac{x + \sqrt{x^2 + 4 q_{p - 4}^2 (1 - x)}}{2}
\right)^{1/2},\ee which appears quite different from the above
$b_{0, q_p} (x)$. From what we learned for the D1/D5 case, we expect
that the two should have the same form if a proper variable is
chosen here. Our experience from the D1/D5 case suggests that  a new
variable $y$ for $b_{q_{p - 4}, 0}$ be chosen here as \be y =
\frac{x + \sqrt{x^2 + 4 q_{p - 4}^2 (1 - x)}}{2} < 1,\ee from which
we can solve to give \be x = \frac{y \left(1 - \frac{q_{p -
4}^2}{y^2}\right)}{1 - \frac{q_{p - 4}^2}{y}}, \quad 1 - x = \frac{1
- y}{1 - \frac{q_{p - 4}^2}{y}}.\ee Note that here $0 < x < 1$ gives
$q_{p - 4} < y < 1$. With these relations, we have \be b_{q_{p - 4},
0} (y) = \frac{y^{\frac{1}{7 - p}}}{7 - p} \left(1 - y\right)^{1/2}
\left(1 - \frac{q^2_{p - 4}}{y^2}\right)^{\frac{p - 5}{2(7 - p)}}
\left(1 - \frac{q_{p - 4}^2}{y}\right)^{-\frac{1}{7 - p}},\ee which
has the exact same functional form as $b_{0, q_p} (x)$ given in
\eqn{pbrane}, as expected. Therefore the delocalized D(p - 4) branes
alone have the same phase structure as the Dp branes. The
non-symmetric appearance for the charges $q_{p - 4}$ and $q_p$ in
either the action \eqn{graction} or the function $b_{q_{q - 4}, q_p}
(x)$ \eqn{gbf} is due to our bias choice of variable $x$ more
favorable to Dp-branes, as in the D1/D5 case. There should exist a
variable $f$, just like the D1/D5 case, with which both $q_{p - 4}$
and $q_p$ appears symmetric in either the action or the function
$b_{q_{q - 4}, q_p}$, given what we know about the phase nature of
either type of branes. From our experience on the D1/D5 case, a good
guess for the present case is \be \label{gfvariable} f =
\frac{\triangle_+}{\triangle_-} = \frac{1 - x}{1 - \frac{q_p^2}{x}}
< 1,\ee which reduces to \eqn{fvariable} when $p = 5$, i.e., the
D1/D5 case. Note that $q_p < x < 1$ gives $0 < f < 1$, where $x =
q_p$ corresponds to $f = 1$ while $x = 1$ to $f = 0$. We solve from
the above action \eqn{gfvariable} for $x$ and have \be \label{gxf} x
= \frac{1 - f + \sqrt{(1 - f)^2 + 4 q_p^2 x}}{2}.\ee In terms of
variable $f$, we have the reduced action \eqn{gractionv}  \bea \bar
I_E &=& - 2\,\bar b \left[\frac{(9 - p) f^{\frac{1}{2}}}{2} +
\frac{7 - p}{2 f^{\frac{1}{2}}} - (8 - p) - \frac{\sqrt{(1 - f)^2 +
4 q_{p - 4}^2 f} + \sqrt{(1 - f)^2 + 4 q_p^2 f}}{2
f^{1/2}}\right]\nn &\,& - \left(1 - f\right)^{\frac{1}{7 - p}}
\left(\frac{1 - f + \sqrt{(1 - f)^2 + 4 q_{p - 4}^2 f}}{2}
\right)^{\frac{1}{2}} \left(\frac{1 - f + \sqrt{(1 - f)^2 + 4 q_p^2
f}}{2}\right)^{\frac{1}{2}},\nn\eea and the function $b_{q_{p - 4},
q_p}$ \eqn{gbf} \be b_{q_{p - 4}, q_p} (f) = \frac{f^{\frac{1}{2}}
(1 - f)^{\frac{p - 6}{7 - p}}}{7 - p} \left(\frac{1 - f + \sqrt{(1 -
f)^2 + 4 q_{p - 4}^2 f}}{2} \right)^{\frac{1}{2}} \left(\frac{1 - f
+ \sqrt{(1 - f)^2 + 4 q_p^2 f}}{2}\right)^{\frac{1}{2}},\ee either
of which is, as expected, symmetric with respect to $q_{p - 4}$ and
$q_p$ and reduces to its correspondence \eqn{ractionf} or \eqn{bf}
when $p = 5$.

The behavior of $b_{q_{p - 4}, q_p} (f)$ vs $f$ will determine the
underlying phase structure when both $q_{p - 4}$ and $q_p$ are
non-zero. Note that we have here $4 \le p \le 6$. For $p = 4$, we
have \be b_{q_0, q_4} (f \rightarrow 0) \rightarrow 0,\quad b_{q_0,
q_4} (f \rightarrow 1) \rightarrow \infty,\ee whose characteristic
behavior when $f \rightarrow 1$ is similar to that when $q_0 = 0$,
i.e., the charged black 4-branes with $\tilde d = 3 > 2$. So we
expect that the present phase structure, though richer and having
details different, is essentially similar to that of charged black
4-branes given in \cite{Lu:2010xt}. To be precise, any given charge
curve crossing the critical line in the two-dimensional charge
square as in the D1/D5 case will give a phase structure similar to
that of charged black 4-branes with $\tilde d = 3$ but in details
more like that of black p-branes with $\tilde d > 3$. The detail
analysis will be similar to the D1/D5 case and will not be repeated
here. For $p = 6$, we have \be b_{q_2, q_6} (f \rightarrow 0)
\rightarrow 0,\quad b_{q_2, q_6} (f \rightarrow 1) \rightarrow q_2
q_6 \ne 0,\ee whose characteristic behavior when $f \rightarrow 1$
is like that of black 5-branes with $\tilde d = 2$. So adding the
delocalized D2 branes to D6 branes modifies the D6 original phase
structure with $\tilde d = 1$\cite{Lu:2010xt} to the one similar to
that of black D5 branes or effectively to that of black p-branes
with $\tilde d > 1$ (Note that though the D5 one is richer, the D5
and D6 phase structures are essentially not much different from that
of chargeless one as discussed in \cite{Lu:2010xt}.). Once again,
the D2/D6 phase structure detail is different and more richer, for
example, having a two-dimensional charge square and the detail
analysis can be similarly performed as for the D1/D5 system and will
not be given here.

\end{document}